\documentstyle[12pt]{article}
\begin{document}
\def\Tr{{\rm Tr}}
\def\tmp{{\tilde m_P}}
\def\MS{M_{SUSY}}
\def\mpl{m_P}

\begin{titlepage}
\begin{center}
June 1992
\hfill LBL-32324 \\
 \hfill UCB-PTH-92/15 \\
\vskip 1.0 in
{\large \bf ANALYSIS OF RUNNING COUPLING CONSTANT UNIFICATION IN STRING
THEORY }\footnote{This work was supported by the
Director, Office of Energy Research, Office of High Energy and Nuclear
Physics, Division of High Energy Physics of the U.S. Department of Energy
under contract DE-AC03-76SF00098 and in part by the National Science
Foundation under grant PHY--90--21139.}\\
\vskip .5 in
{
{\bf Mary K. Gaillard and Rulin Xiu} \\
\vskip 0.5 cm

{\it Department of Physics, University of California\\ and\\
Physics Division, Lawrence Berkeley Laboratory, 1 Cyclotron Road\\
Berkeley, CA 94720 \\ } }
\end{center} \vskip 1.0 in \begin{abstract}
We use recently obtained 2-loop string coupling constants to
analyze a class of string models based on orbifold compactification.
Assuming weak coupling at the string scale $m_s$ and single-scale
unification leads to restrictions on the spectrum of massive ($M_Z\ll M\le
m_s$) matter supermultiplets and/or on the Kac-Moody algebra level.
\end{abstract}
\end{titlepage}

\pagestyle{empty}
\newpage

\pagestyle{plain}
\pagenumbering{arabic}

 In the past few years, progress in string phenomenology has been very
encouraging. On one hand, the elucidation of the effective theories that
describe the low energy dynamics of the light string excitations has been
carried out to one genus and higher genus~\cite{dkl},\cite{higher} in string
calculations.
A
systematic scheme has been developed to obtain an effective theory with modular
invariance to all loop orders. One loop and all loop gauge
couplings in string effective theories have been obtained for a class of models
\cite{mary}--\cite{linear}. On the other hand, the
construction of ``realistic'' string models has been
developed to produce some
explicit string models. Even though none of them are totally realistic, some
are very close to the
standard model and might even make some good predictions
\cite{font}. Because of this progress, it is now possible confront specific
models with data.

 In this paper, we discuss the running gauge coupling constant unification in
string effective theories. It is interesting to note that string unification
has, in general, a more flexible meaning than standard model unification. On
the other hand, because of the relation between the unification scale and the
gauge coupling constant in string effective theory, the string analysis is
more restrictive. We restrict our discussion to the minimal string model from
an $N=1$ orbifold with no $N=2$ space-time supersymmetry. In this case, the
running coupling constants do not have nontrivial moduli-dependent string
threshold corrections~\cite{higher}. By minimal string model, we mean that the
model has \( SU(3) \times SU(2) \times U(1) \) as observable gauge group (that
is, we neglect possible mixing with additional $U(1)$'s, and do not consider
intermediate scale gauge symmetry breaking) and N=1 supersymmetry below the
compactification scale. In addition to the standard model matter content there
are generally extra particles which become massive at some high energy scales.
Our analysis will impose restrictions on the spectra of massive states. We pay
particular attention to this kind of model not only because it is simple to
analyse but also because it is the closest approximation to the standard model
to date.

In the following we give the two-loop renormalization group
equations (RGE) in string effective theory for a class of models
derived from orbifold compactification. We take as a
starting point the result of~\cite{mary} that suggests that two-loop
unification occurs at the string scale $m_s$. We will see that, under the
assumption of weak coupling at the string scale, the scale $\tmp = {1\over 2}
m_Pe^{1\over 2}$, where $\mpl$ is the Planck mass, can be taken as the
effective string unification scale in the first order RGE.

 Whereas the unification of the running gauge coupling constants at some scale
is a coincidence in the standard model, a relation among gravitational and
gauge coupling constants at the string scale is required in string theory. It
has been shown \cite{gins} that at the string scale \( m_{s}^{2}/m_P^{2} =
\alpha_{i}k_{i}, \;\;m_{s} = 2/\sqrt{\alpha'}\). Here the Planck mass $m_P$ is
defined by \( G_{N} = \kappa^{2}/8\pi = 1/m_P^{2}\) . \( \alpha' \) is the
string tension and $k_{i}$ is the Kac-Moody level of the gauge algebra \( G_{i}
\). For a nonabelian gauge group its level is an integer and is determined by
the corresponding Kac-Moody algebra realized on a world sheet. For a U(1) gauge
group the level is defined to be Y = \( k_{1}/2 \) \cite{level}, where Y is the
the double-pole coefficient in the operator product expansion of the
corresponding Kac-Moody currents. The relation is obtained under the assumption
that any gauge symmetry has its origin in an internal Kac-Moody algebra as is
the case for all known closed string theories. It is a general result in the
sense that it does not depend on the particular realization of a gauge algebra
in a model. To see the dependence of coupling constants on a scale, a string
loop calculation has to be applied.  In the following analysis,
the $U(1)$ charge is the
weak hypercharge of the standard model, but normalized as in $SU(5)$, so that
$k_1=k_2=k_3$ corresponds to coupling constant unification in the usual sense.

In string effective theory, the gauge coupling constant $g$ is determined by
the vacuum expectation value of the dilaton $s$: \( g^{-2} = < Re s > \). The
one loop modular invariant gauge coupling has been calculated \cite{mary} for
orbifolds with gauge group \( E_{8}\bigotimes {E_{6}\bigotimes{U(1)^{2}}} \)
and $k_a=1$. A correspondence between the effective coupling constant and the
running coupling constant in quantum field theory was proposed in~\cite{mary},
where it was shown that the relation can be interpreted as the two loop running
coupling constant~\cite{susy} with the string scale \( m_{s}\) as the two loop
unification scale, up to possible string threshold corrections. More precisely,
the results of~\cite{mary} suggest the identification $g^2_a({1\over
2}g^2M_P^2) = g^2/k_a - \epsilon_a/4\pi$, where $M_P = m_P/\sqrt{8\pi}$ is the
reduced Planck mass, and $\epsilon_a$ is a small threshold correction. Here
the running coupling constant $g^2(\lambda^2)$ is to be interpreted as the
measured coupling constant with external momenta $-p^2 = \lambda^2$. This is
related~\cite{tatu} to the scale $\mu$ of the $\overline{MS}$ regularization
scheme by $\lambda^2 = e\mu^2$. In this scheme the unification scale becomes
$\mu_s^2 = e^{-1}\alpha m^2_P/4 = \alpha\tmp^2.$ Consequently, for scales
$\mu>M_{SUSY}$, the running gauge coupling constants in string phenomenology
can be written as: \begin{eqnarray} g_a^{-2}(\mu) = g^{-2}_a - {C_a^G\over
8\pi^2}\ln2 - \frac{1}{8 \pi^{2}} ( 3C_{a}^{G} - C_{a}^{M}) \ln
\frac{\mu_s}{\mu} + \frac{C_{a}^{G}}{8\pi^{2}} \ln
\frac{g_a^{2}}{g_a^{2}(\mu)}\nonumber\\ + \frac{1}{8 \pi^{2}} \sum_{A} C_a^{A}
\ln \frac{Z_A(\mu_s)} {Z_A(\mu)} - \frac{1}{2}\sum_{I} \beta_{a}^{I} \ln[
\eta^{2}{(it^{I}) }\overline{\eta^{2}(it^I)}(t+\bar{t})^{I}] .\end{eqnarray}
Here $t^I$ are the moduli and \(4e \mu_s^{2}/m_P^{2} = \alpha_{a}k_{a} = \alpha
\). The index $A$ denotes the matter content: $C^M_a = \sum_A C^G_a$. $C^G_a$
is the quadratic Casimir operator in the adjoint representation of the gauge
subgroup labeled by $a$, and $C^A_a = \Tr(T^A_a)^2$, where $T^A_a$ is any
generator of the gauge group $G_a$ for the representation $R_A$, is related to
the quadratic Casimir operator $C_2^a(R_A)$ by dim$(G_a)C^A_a =
$dim$(R_A)C_2^a(R_A).$ The matter wave-function normalization satisfies at one
loop \begin{eqnarray} \frac{d\ln Z_{A}(\mu)} {d\ln \mu} =
\sum_bC_{2}^b(R_{A})\frac{\alpha_b}{\pi}. \end{eqnarray} Now we write eq.(1)
as: \begin{equation} \alpha^{-1}_a(\mu)= k_{a}\alpha^{-1} + \frac{b_{a}}{2
\pi} \ln \frac{ \tmp}{\mu} + \delta_{a}^{(2)} + \Delta_{a}, \;\;\;\; b_a =
C^M_a - 3C^G_a. \end{equation} Here \(\delta_{a}^{(2)}\) is the ``second
order'' correction; integration of (2) using the one-loop expression for
$\alpha(\mu)$ gives, for scales $\mu>M_{SUSY}$: \begin{equation}
\delta_{a}^{(2)} = \frac{b_{a}} {2 \pi} \ln \alpha - \epsilon_a +
{\frac{C_{a}^{G}} {2\pi}}\ln{\frac{\alpha_a}{\alpha_a(\mu)}} +
\sum_{A,b}\frac{1}{\pi b_b} C^A_{a}C^b_2(R_A)\ln\frac{\alpha_b(\mu_s)}
{\alpha_b(\mu)} , \end{equation} where $\epsilon_a = C_a^G\ln 2/2\pi$, and \(
\Delta_{a}\) is the string threshold correction~\cite{dkl}: \begin{equation}
\Delta_{a} = - 2\pi\sum_{I} \beta_{a}^{I} \ln[ \eta^{2}(it^{I})
\overline{\eta^{2}(it^{I})}(t+\bar{t})^{I}]. \end{equation} Eq.(3) expresses
the result that in first order the running coupling constants approximately
``unify '' at Planck scale with some string threshold corrections.

Although the result (1) was explicitly derived~\cite{mary} for the \(
E_{8}\bigotimes {E_{6}\bigotimes{U(1)^{2}}} \) orbifold, its generalization to
other $Z_N$ orbifolds with $k_a = 1$ is straightforward. We will also consider
the assumption that it can be extended to $k_a \ne 1$. Here we restrict our
analysis to orbifolds with no $N=2$ supersymmetry sectors; in this
case~\cite{higher} \begin{equation} \beta_a^I = 0 = \Delta_a. \end{equation}
RGE analyses including moduli-dependent threshold corrections have been
considered elsewhere~\cite{mod}. The restriction (6) implies a
condition~\cite{dkl}--\cite{mary} on the spectrum of gauge nonsinglets and
their modular weights.

For the minimal string model, the above results then give.
\begin{equation}
 \alpha_a^{-1}(M_{Z}) = k_{a}\alpha^{-1} + \frac{b_{a}^{(I)}}{2 \pi} \ln
\frac{M_{SUSY}}{M_{Z}} +
 \frac{b_{a}^{(II)}}{2 \pi} \ln \frac{\tmp}{M_{SUSY}} + \Pi_{a}
 + \delta_{a}^{(2)} .
\end{equation}
Here
\begin{equation}
\Pi_{a} = \frac{1}{2\pi}\sum_{E}b_{aE}\left( 1 +
 \frac{\alpha_{a}}{2\pi}C^G_{a} +
 \sum_b\frac{\alpha_{b}}{\pi}C^E_{a}C_{b}(R_E)\right)\ln\frac{\mu_s}{m_{E}},
\end{equation}
where we have used the first order expression for $\ln[\alpha_a(\mu_s)
/\alpha(m_E)]$ in the second order terms.
The index $E$ extends over extra, massive particles.
 \[ b_{a}^{(I)} = \left( \begin{array}{c} 4.2\\ -3\\ -7 \end{array}
\right) \] is the $\beta$-function after supersymmetry is broken (we assume two
Higgs doublets with mass $m_H\approx M_Z$), and \[
b_{a}^{(II)} = \left( \begin{array}{c} 6.6\\ 1\\ -3 \end{array}
\right) \] is the $\beta$-function in the supersymmetric region for the minimal
supersymmetric standard model (MSSM).

Note that in string phenomenology, the parameters \( \alpha \), \( M_{SUSY} \),
\( \ln (\mu_s/m_{E}) \), \( M_{W} \) are supposed to be determined dynamically.
Then the coupling constants at the weak scale can in principle be computed. In
addition, according to the most popular supersymmetry breaking mechanism,
gaugino condensation~\cite{nilles}, \(M_{SUSY}\) is determined by the gauge
group in the hidden sector. Then eq.(7) implies that low energy measurements
can in principle put constraints on both the hidden sector gauge group and the
observable gauge group. Unfortunately, there is not as yet a satisfactory
theory of supersymmetry breaking by hidden gaugino condensation; we will use
the phenomenological constraint that $\MS$ must be within about an order of
magnitude of a $TeV$, and define $M_T = 1TeV$ as the ``first order'' SUSY
breaking scale.

We first consider the case $k_a=1;$ presently most string models produced are
of this kind. We note that the result (3) was obtained~\cite{mary} from the
analysis of the one-loop chiral anomaly, which is connected by supersymmetry to
the conformal anomaly, in the context of supergravity. In the context of
supersymmetric Yang-Mills theory, the authors of Ref.~\cite{susy} argue that
the two-loop condition (3) should be valid to all loop orders; their argument
is related to the well-known result that the chiral anomaly is determined
completely at the one loop level. We therefore minimize the approximations in
the evaluation of (3); specifically, we do not use the lowest order expression
for $\ln[\alpha(\mu_s) /\alpha(\MS)]$. In addition to the one-loop
approximation used in writing the last term in (4), we make the following
approximations. We use the lowest order expression for
$\ln[\alpha(\MS)/\alpha(M_W)]$ to evaluate $\alpha(\MS)$. This gives, using LEP
data~\cite{data} and other results~\cite{lang}: \begin{equation}
\alpha^{-1}(\MS) + b_{a}'^{(I)}z = {\tilde B}_{a}, \;\;\;\; z = \frac{1 } {2
\pi} \ln \frac{M_{SUSY}}{M_{T}}, \end{equation} where the quantities \[ {\tilde
B} = \left( \begin{array}{c} 57.25\pm .11\\ 30.84\pm .11\\ 11.23\pm .80
\end{array}\right), \;\;\;\; b_{a}'^{(I)} = \left( \begin{array}{c} 4.30\\
-2.86\\ -7.23 \end{array}\right), \] include the two-loop corrections as
obtained in this approximation. We used as input the values quoted
in~\cite{lang}, but with a larger error ($\delta\alpha_3^{-1}= \pm .79$) on
$\alpha_s$ which covers the slightly higher recent result~\cite{data3}. We then
evaluate (3) taking $\mu = \MS$, and approximating $\ln\alpha_a(\MS) =
-\ln{\tilde B}_a+{\tilde B}_a^{-1}b'^{(I)}_az$, which is justified by the
assumption $.1< \MS/M_T < 10,\;\; |z|\le .37$, and $\ln\alpha_a(\mu_s) =
\ln\alpha$, which is justified (for $k_a=1$) under our weak coupling
assumption: $\alpha^{-1}\gg \epsilon_a = (0,.22,.33)$. This gives:
\begin{equation} \alpha^{-1} + \frac{b_{a}'^{(I)} - b_{a}^{(II)}}{2 \pi} \ln
\frac{M_{SUSY}}{M_{T}} + \Pi_{a} = B_{a} + c_a\ln\alpha , \end{equation} \[ B =
\left( \begin{array}{c} 18.87\pm .11 +.03\eta\\ 19.98\pm .11+.04\eta\\ 27.01
+.82\eta \end{array}\right), \;\;\;\; c_{a} = \left( \begin{array}{c} 1.11\\
1.53\\ -.106 \end{array}\right),\] where $\eta= \pm 1$ is the sign of
$\delta\alpha_3^{-1}$.

We are interested in finding constraints on string phenomenology
from low energy measurements. We write (9) in the form:
\begin{equation}
 \alpha^{-1} - 1.11\ln\alpha - 2.57 z + \Pi_{1} = B_{1},
\end{equation}
\begin{equation}
 \alpha^{-1} - 1.53\ln\alpha - 4.09 z + \Pi_{2} = B_{2},
\end{equation}
 \begin{equation}
 \alpha^{-1} + .106\ln\alpha - 4.40 z + \Pi_{3} = B_{3}.
\end{equation}
These equations have no solution for $\Pi_a=0$. In this case the analysis
reduces to that of MSSM which has been
 done \cite{data1} to second order and gives: \( M_{GUT} = 10^{16\pm0.3}\)GeV
 and \( \alpha_{GUT} \simeq 26\). In string coupling constant
 unification, we have an additional constraint:
 \( \alpha_{GUT} m_P = M_{GUT} = \mu_s \), which cannot be satisfied.

However, the universal anomaly cancellation condition (6) generally requires
the existence of additional particles, so we expect $\Pi_a\ne 0$. Since
$\Pi_a\ge 0$, if we impose the constraint $|z|\le .37$,
each of eqs. (11)--(13) gives an upper bound on $\alpha^{-1}$; the
strongest bound is from (11):
\begin{equation}
\alpha^{-1}\le 16.83
\end{equation} (Note that although the first
term on the right hand side of (4) is not strictly speaking of second order,
this bound, together with the weak field assumption $\alpha\le 1$ assures that
it, as well as the threshold correction $\epsilon_a$, is comparable to the
second order corrections.)

Next consider the difference
\begin{equation}
\Pi_3 - \Pi_2 = 7.03 -1.64\ln\alpha + .31z\pm .11 + .78\eta\ge 6.03
\end{equation}
using the weak coupling constraint $\alpha\le 1$.
This tells us that for this kind of model to work, at least one extra SU(3)
triplet is required. Note that (15) is independent of $k_1$, so our analysis
excludes the 3-generation \( SU(3) \times SU(2) \times U(1) \) model of
\cite{font}, which is a level one model with no extra SU(3) triplets.
For general orbifold compactifications based on the heterotic string, the
chiral multiplets in the observable sector are in $27 + {\overline{27}}$ of
$E_6$. The massive states come in particle-anti-particle pairs that we denote
according to their $(SU(3)_c\times SU(2)_L)_{|Q|}$ quantum numbers ($Q$=
electric charge) as
$$ q = (3,2),\;\;\;\; \ell = (1,2), \;\;\;\; d = (3,1)_{1\over 3},\;\;\;\;
u = (3,1)_{2\over 3},\;\;\;\; e=(1,1)_1.$$
Then
$$ \Pi_a = \sum_ib_a^i\Pi_i, \;\;\;\; \Pi_i = \sum_{M_i}{n_i\over \pi}
\ln{\mu_s\over M_i}\ge 0,$$ where $n_i$ is the number of pairs with
mass $M_i$. Using the first order expressions for the $\beta$-function
constants $b_a^i$, Eq.(8), gives
$$\Pi_1 = .1\Pi_q + .8\Pi_u + .2\Pi_d + .3\Pi_{\ell} + .6\Pi_e, $$
\begin{equation}
\Pi_2 = 1.5\Pi_q + .5\Pi_{\ell},\;\;\;\;
\Pi_3 = \Pi_q + .5\Pi_u + .5\Pi_d.
\end{equation}
Writing
$$\Pi_1 -.4\Pi_3 + .2\Pi_2 = 12.06 -.8\alpha^{-1} + 1.46\ln\alpha +
1.63z\pm .13 -.29\eta
$$ \begin{equation}
= .6\Pi_u + .4\Pi_{\ell} + .6\Pi_e\ge 0,
\end{equation}
gives a stronger bound on $\alpha$:
\begin{equation}
\alpha^{-1}\le 11.84.
\end{equation}
Using the weak coupling constraint,
we can also exclude models in which the heavy states all transform as
$(3,1) + (\bar{3},1) + (1,2)$:
$$\Pi_1 -.4\Pi_3 - .6\Pi_2 = - 1.2\Pi_q + .6\Pi_u + .6\Pi_e $$
\begin{equation}
= -3.92 + .23\ln\alpha - 1.64z\pm .18-.32\eta \le - 2.81,
\;\;\;\; \Pi_q\ge 2.34,
\end{equation}
for $\alpha\le 1$. In addition we get, using (18),
$$\Pi_3 -{2\over 3}\Pi_2 = .5\left(\Pi_u + \Pi_d -{2\over 3}\Pi_{\ell}\right)
$$ \begin{equation}
= 13.69 -{1\over 3}\alpha^{-1} - 1.13\ln\alpha \pm .07
+ .79\eta + 1.67z \;\;\cases{ \ge 11.06\cr \le 15.42}.
\end{equation}
which in turn implies that $(3,2)+({\bar{3}},2)$ states do not make the full
contribution to $\Pi_3$. Using (18) and $\alpha\le 1$, together with the above
results, gives
$$\Pi_2 \le 20.54,\;\;\;\; \Pi_3 \le 28.41,$$
\begin{equation}
2.34\le \Pi_q \le 13.69,\;\;\;\; 22.12\le \Pi_d + \Pi_u \le 52.14.
\end{equation}
The above bounds have implications for the masses of these states; assuming
that $\Pi_q$ is dominated by the lightest multiplet, we find, for example
\begin{equation}
1.3 GeV \le m_q\le 3.9\times 10^{15}GeV.
\end{equation}
Clearly the lower mass range is ruled out; if the upper bound on $\Pi_q$
is saturated, we require $n_q \ge 2$, which in turn implies $m_q\ge 2.8\times
10^9GeV$. The value found for $\Pi_d + \Pi_u$ is has to be
interpreted in terms of a lower bound on the number of these states: $n_u+n_d
\ge 2$. If the upper bound on $\Pi_d + \Pi_u$ is saturated, we require
$n_u+n_d\ge 5$; the corresponding mass bounds are
\begin{equation}
m_{u\;{\rm or}\;d}\le 5TeV, \;\;{\rm if}\;\; n=2, \;\;\;\;
m_{u\;{\rm or}\;d}\ge 36TeV \;\;{\rm if}\;\;n\ge 5.
\end{equation}

Finally, we consider models with arbitrary levels: $k_a\ge 1$.
Although no minimal string model with a higher Kac-Moody level has as yet
been constructed, the basic technique has already been
established~\cite{level}, and there
seems to be no fundamental difficulty in obtaining such a model. If the low
energy effective theory differs from the $k_a = 1$ case only by the additional
factors $k$ in the couplings of the gauge supermultiplets to the dilaton $s$,
then eq.(3) should still be valid. There is an additional threshold-like effect
from the last term in (4): $\ln\alpha_b(\mu_s) = \ln\alpha -\ln k_b$.

If the model is clever enough, it may be possible to achieve universal
anomaly cancellation without extra particles. We will only consider this
more restrictive case; then the low energy
measurements constrain the values of \( k_{1}, k_{2}\) and \( k_{3}\).
First we note that for models that unify to a larger gauge group at the string
scale, \( \alpha_{1}^{-1} = \alpha_{2}^{-1} = \alpha_{3}^{-1}
= k\alpha^{-1} \), we would have the constraint:
 \( M_{GUT} = \mu_s = k\alpha_{GUT} m_P \ge \alpha_{GUT} m_P \),
which cannot be satisfied without additional particles or nontrivial string
threshold corrections or both. For the models we are considering, which do
not have nontrivial string threshold corrections, extra $SU(3)$ triplets are
again required for a model to work.

Setting $\Pi_a= 0$ eqs.(11)--(13) now appear as:
\begin{equation}
 k_{1}\alpha^{-1} - 1.11\ln\alpha- 2.56 z = B_{1}(k),
\end{equation}
\begin{equation}
 k_{2} \alpha^{-1} - 1.53\ln\alpha - 4.09z = B_{2}(k),
\end{equation}
\begin{equation}
 k_{3} \alpha^{-1} + .106\ln\alpha - 4.41 z = B_{3}(k),
\end{equation}
\[ B_a(k) = B_a + \sum_{A,b}\frac{1}{\pi b_b} C^A_{a}C^b_2(R_A)\ln k_b.\]
If $\ln k_a \sim 1$, the additional correction is of same order as the
uncertainty in the SUSY threshold.
Using the constraints $-.37\le \alpha\ln\alpha\le
0$ for $0\le\alpha\le 1$ and $|z|\le .37$, (25) and (26) give
\begin{equation}
1.16\le {k_3\over k_2}\le 2.32,
\end{equation}
if we restrict the integers $1\le k_2,k_3\le 10.$ This allows for the
possibility $k_2 = 2,\;k_3 =1$;
however, the solution to (25) and (26) with these values gives
$9.5\le\alpha^{-1}\le 12$, $M_{SUSY}\le 3GeV$, which is excluded.
For fixed $k_3/k_2=2$ and $k_2\alpha^{-1}\ge 4.2$, $z$ decreases as $k_2$
increases, so there is no solution
with $k_3/k_2 = 2$; in a similar way one can exclude the cases $k_3/k_2 = 7/4,
\;5/3\;,8/5,\ldots.$ We find
solutions with $k_2 = 5,\; k_3 = 7$, $3\le\alpha^{-1}\le 4.6$,
$4.1\le k_1 \le 4.7$, $-1.3\le z\le .8$, and with $k_2 = 6,\; k_3 = 8$,
$3\le\alpha^{-1}\le 4.7$,
$4.8\le k_1 \le 5.3$, $-.59\le z\le 2$, which cover the allowed range for
$M_{SUSY}$, and are the only solutions with $k_3/k_2 = 7/4$ and $4/3$,
respectively. In addition we find that
$k_2 = 2,\; k_3 = 3$, $ \alpha^{-1} \approx 9.3,\;k_1 \approx 1.8$, and $k_2 =
4,\; k_3 = 6$, $\alpha^{-1}\approx 4.5,\;k_1\approx 4$, are allowed within the
errors but with marginally acceptable values for $z$: $M_{SUSY}\ge 7.15TeV$
in the first case and $M_{SUSY}\le 140GeV$ in the second case; these are the
only solutions with $k_3/k_2 = 3/2$.

True coupling constant unification with $k_a =1$ is of course the most
attractive possibility. The relation between the coupling and the scale at
unification in orbifold compactification is inconsistent with the values
determined~\cite{data1} from a fit of the MSSM to the data. This could be
evaded without introducing extra particles in several ways. In Calabi-Yau
compactification, it is generally assumed that the unification scale is
determined by the vacuum expectation value of a modulus field $t$ rather than
the dilaton $s$, in which case the relation between the coupling constant and
the unification scale would not hold. Since these theories have not been
studied to the same extent as orbifold models, no conclusion can really be
drawn. Another possibility is that the MSSM is valid to the scale of
$10^{16}GeV$, above which there is a larger gauge group; in this case nothing
can be inferred from the data on the value of $\alpha = m^2_s/\mpl^2$. A third
alternative is orbifold models with moduli-dependent threshold
corrections~\cite{mod}. However in theses models, the moduli-dependence of the
effective potential for gaugino condensation~\cite{pot} breaks the continuous
symmetries that would otherwise protect~\cite{bg} scalar particles from
acquiring large masses when local supersymmetry is broken. For this reason,
models with universal anomaly cancellation may be more attractive for
phenomenology. We have seen that these models can fit the data, which in turn
provide restrictions on the spectrum of heavy gauge nonsinglet
supermultiplets; in particular, there must be color triplet (3,1) and (3,2)
states. Some of these could be light enough to be produced in collider
experiments, and, for a suitable mass range, these states could have
interesting implications for proton stability and/or baryogenesis. A similar
analysis including moduli-dependent threshold corrections and extended gauge
groups will be given elsewhere.

\vskip 28pt
\noindent{\bf Acknowledgement.}
\vskip 12pt
This work was supported in part by the
Director, Office of Energy Research, Office of High Energy and Nuclear Physics,
Division of High Energy Physics of the U.S. Department of Energy under Contract
DE-AC03-76SF00098 and in part by the National Science Foundation under grant
PHY-90-21139.

\end{document}